\documentclass[5p,twocolumn,times,number]{elsart5p}
\usepackage{ae}
\usepackage{amsmath}
\usepackage{amssymb}
\usepackage{graphics}
\usepackage{graphicx}
\usepackage{epsfig}
\usepackage{dcolumn}
\usepackage{bm}

\begin{document}
\begin{frontmatter}

\title{Study of timing properties of single gap high-resistive bakelite RPC}
\author{S. Biswas\thanksref{label1}\corauthref{cor1}},
\ead{saikatb@veccal.ernet.in}
\thanks[label1]{}
\corauth[cor1]{}
\author[label2]{S. Bhattacharya},
\author[label2]{S. Bose},
\author[label1]{S. Chattopadhyay},
\author[label2]{S. Saha},
\author[label3]{Y.P. Viyogi}

\address[label1]{Variable Energy Cyclotron Centre, 1/AF Bidhan
Nagar, Kolkata-700 064, India}
\address[label2]{Saha Institute of Nuclear Physics, 1/AF Bidhan Nagar, Kolkata-700
064, India}
\address[label3]{Institute of Physics, Sachivalaya Marg, Bhubaneswar, Orissa-751 005, India}

\begin{abstract}
The time resolution for several single gap (2 mm) prototype
Resistive Plate Chambers (RPC) made of high resistive
($\rho$ $\sim$ 10$^{10}$ - 10$^{12}$ $\Omega$ cm), 2 mm thick matt finished
bakelite paper laminates with silicone coating on the inner surfaces,
has been measured. The time resolution for all the modules has been found to be $\sim$ 2 ns
at the plateau region.

\end{abstract}
\end{frontmatter}

\section{Introduction}
\label{}
Performances of several single gap (2 mm) prototype Resistive Plate Chambers (RPC) \cite{RSRC81}
made of high resistive ($\rho$ $\sim$ 10$^{10}$ - 10$^{12}$ $\Omega$ cm)
bakelite paper laminates
produced and commercially available in India has been carried out in recent times \cite{SB109}. A thin silicone coating
has been applied to the inner electrode faces of the detectors to make the surfaces smooth. Such
high resistive electrodes are being explored since the detectors are
one of the candidates of the proposed neutrino oscillation experiment
in the India-based Neutrino Observatory (INO) \cite{INO06}. One of the requirements for
the INO RPC is to have a time resolution $\sim$ 2 ns or better.

The silicone coated chambers, operated in the streamer mode using
argon, tetrafluroethane (R-134a) and isobutane in 34:59:7 mixing ratio,
prepared by a gas mixing and flow control unit \cite{SBose109}
have been tested with cosmic rays. The results of the long
term test (with efficiency $>$ 90\%, counting rate of $\sim$ 0.1 Hz/cm$^2$)
and some other aspects such as
crosstalk, dependence on threshold value, the effect of external
humidity etc. of these silicone coated RPCs have been reported
earlier \cite{SB109,SB209}. In this article, we would like to present
the timing characteristics of such single gap (2 mm) silicone coated RPCs.

\section{Test setup and method of calculation}
\label{}

\begin{figure}
\includegraphics[scale=0.62]{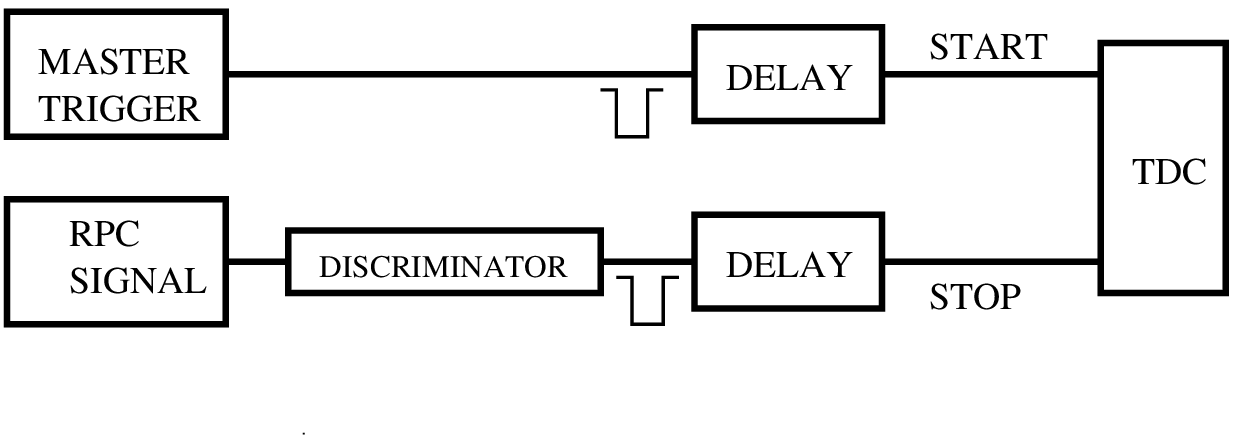}\\
\caption{\label{fig:epsart}Schematic representation of the time resolution
measurement setup. A common start
Phillips Scientific 7186 Time to Digital Converter (TDC) was used.} \label{fig:1}
\end{figure}
The time resolution of the RPC was measured  in the same cosmic ray
test bench described in Ref. \cite{SB109}. The cosmic ray telescope was
constructed using three scintillators, two placed above the RPC and
one below. The individual time resolution of each RPC was estimated
as follows. The triple coincidence of the signals
obtained from the three scintillators was taken as the START signal (master
trigger) for the TDC. The STOP signal was taken from a single RPC strip.
Fig.~1 shows the schematic of the time resolution measurement setup.

\begin{figure}
\includegraphics[scale=0.32]{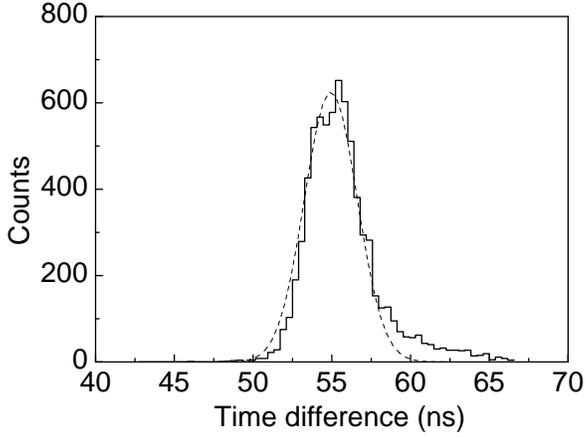}\\
\caption{\label{fig:epsart} The distribution of the time difference
between the RPC and the master trigger.}\label{fig:2}
\end{figure}

The distribution of the time difference between the
master trigger and the signal from one RPC strip is shown in Fig.~2.
Time calibration was measured as 0.1 ns/channel. From the time difference
spectrum, the full width at half maximum (FWHM) and the corresponding standard deviation
($\sigma$$_{ij}$), where $i$ and $j$ refer to scintillators and the RPC, were obtained by
fitting a Gaussian function. The same $\sigma$$_{ij}$ were obtained similarly for the
3 different pairs of the scintillators I, II and III. The intrinsic time resolutions of the
RPC and the scintillators were obtained from the individual standard deviations $\sigma$$_{i}$,
$\sigma$$_{j}$, which were extracted by solving the equations:
$\sigma$$_{ij}$$^{2}$ = $\sigma$$_{i}$$^{2}$+$\sigma$$_{j}$$^{2}$ \cite{PF}.
Time resolution (FWHM) of the individual scintillators were obtained as: 3.20 $\pm$ 0.07 ns (scintillator I),
3.39 $\pm$ 0.08 ns (scintillator II) and 1.98 $\pm$ 0.02 ns (scintillator III), where the quoted uncertainties
include statistical and fitting errors. The extracted time resolution (FWHM) of the RPC at 8 kV operating
voltage for a typical run was: 2.48 $\pm$ 0.08 ns.

\section{Results}
\label{}
\begin{figure}
\includegraphics[scale=0.32]{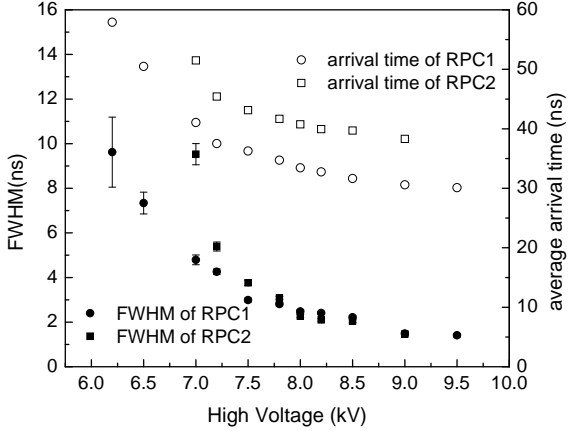}\\
\caption{\label{fig:epsart} The time resolution (FWHM) and the
average signal arrival time with respect to the master trigger as a
function of HV for two silicone coated RPC.}\label{fig:3}
\end{figure}

The average signal arrival time (T), taken as the mean position of the
fitted Gaussian peak in the time difference spectrum (such as in Fig.~2)
and the time resolution (FWHM) $\tau$ of the RPC as function of the applied high voltage (HV) for the two
30 cm $\times$ 30 cm RPCs are shown in Fig.~3. The measured values of T include the delays introduced
in both the START and STOP channels by the electronics shown in Fig.~1.
The time resolution for both the modules
improves and the average signal arrival time decreases with the
increase of HV which is common to any gas filled
detector. At the plateau region, the time resolution has been found
to be $\sim$ 2 ns.

One of the modules was tested for a long period (more than 130 days) at a
constant high voltage of 8 kV and showed nearly constant values of the time resolution ($\sim$ 2-3 ns)
and the average signal arrival time ($\sim$ 55 ns) as shown in Fig.~4.

\begin{figure}
\includegraphics[scale=0.32]{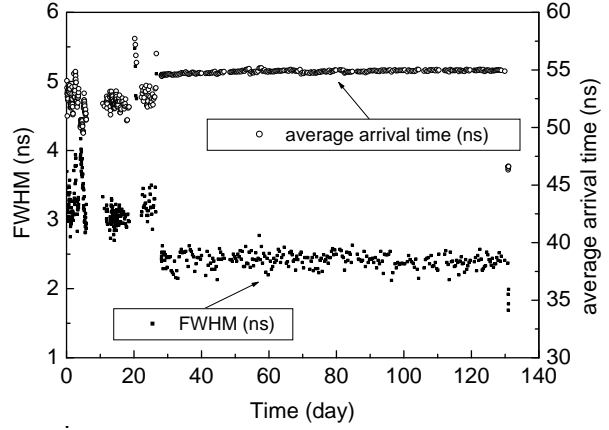}\\
\caption{\label{fig:epsart}The time resolution (FWHM) and the average signal
arrival time as a function of period of
operation.}\label{fig:4}
\end{figure}

\label{}
\begin{figure}
\includegraphics[scale=0.32]{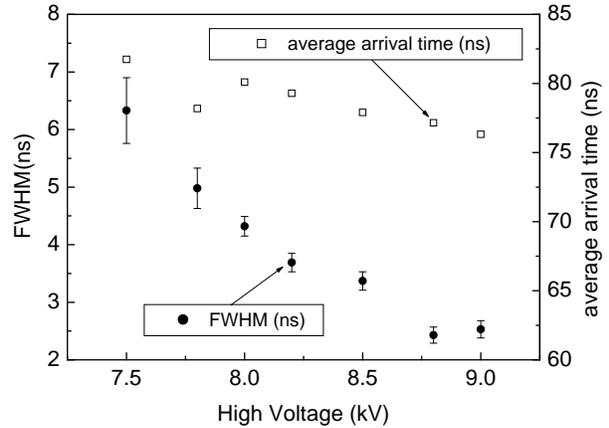}\\
\caption{\label{fig:epsart} The time resolution (FWHM) and the
average signal arrival time of one RPC with respect to another RPC as a
function of HV.}\label{fig:5}
\end{figure}

Finally the time resolution between two RPCs was measured by taking the START
signal from one RPC operated at constant voltage (8 kV),
while the voltage of the other RPC was varied. The results are shown in Fig.~5. In
this case, the average time resolution is found to be $\sim$ 3 ns in the plateau region.

\section{Conclusions and outlook}
\label{}
In conclusion, a systematic study on the timing properties of silicone
coated RPCs made of bakelite paper laminates, commercially
available in India has been performed. The measured time resolution
of those RPCs have been found to be $\sim$ 2 ns which is comparable
to any single gap glass or linseed oil coated bakelite RPC. The
study of the effect of continuous HV on the time resolution has also been performed.
In the long term operation of the detector, 2-3 ns time resolution is obtained.

\section{Acknowledgement}
\label{}
We would like to thank Mr. Ganesh Das of VECC for
fabricating the detectors.\\

\end{document}